\documentclass[useAMS,usenatbib]{mn2e}
\usepackage{rotating}
\usepackage{graphicx}
\usepackage{color}
\usepackage{epsf}
\usepackage{amsmath}
\usepackage{aas_macros}
\newcommand{\msun}{M_{\odot }}
\newcommand{\Mo}{\msun}
\newcommand{\Ro}{R_{\odot }}

\newcommand{\ud}{{\rm d}}
\newcommand{\s}{{\rm s}}

\newcommand{\BSE}{{\sc BSE}}

\title[LGRBs in Massive Binaries]
{
The properties of long gamma-ray bursts in massive compact binaries
}

\author[R.P.~Church et al.]{Ross P.~Church$^{1}$\thanks{email: ross@astro.lu.se},
Chunglee Kim$^{2,1}$, Andrew J. Levan$^{3}$ and Melvyn B. Davies$^{1}$\\
$^{1}$Lund Observatory, Department of Astronomy and Theoretical Physics, Box 43, SE--221 00, Lund, Sweden.\\
$^{2}$Department of Physics, West Virginia University, Morgantown, WV 26506, USA.\\
$^{3}$Department of Physics, University of Warwick, Coventry, CV4 7AL \\
}

\begin{document}

\date{Accepted 2012 XXXber XX . Received 2012 XXXber XX; in original form 2012
 October XX}

\pagerange{\pageref{firstpage}--\pageref{lastpage}} \pubyear{2012}

\maketitle

\label{firstpage}

\begin{abstract}

We consider a popular model for long-duration gamma-ray bursts, in which the
progenitor star, a stripped helium core, is spun up by tidal interactions with a
black-hole companion in a compact binary.  We perform population synthesis
calculations to produce a representative sample of such binaries, and model the
effect that the companion has on material that falls back on to the newly-formed
black hole.  Taking the results of hydrodynamic models of black-hole formation
by fallback as our starting point, we show that the companion has two principal
effects on the fallback process.  First, a break forms in the accretion curve at
around $10^4\,\s$.  Secondly, subsequent to the break, we expect to see a flare
of total energy around $10^{50}\,{\rm erg}$.  We show that the break and flare
times are set largely by the semi-major axis of the binary at the time of
explosion, and that this correlates negatively with the flare energy.  Although
comparison with observations is non-trivial, we show that our predicted break
times are comparable to those found in the X-ray light curves of canonical
long-duration gamma-ray bursts.  Similarly, the flare properties that we predict
are consistent with the late-time flares observed in a sub-sample of bursts.

\end{abstract}


\section{Introduction}

The launch of the {\it Swift} satellite in 2004 has provided a massive step
forwards in our ability to observe gamma-ray bursts, in particular at early
times \citep{Gehrels04}.  The combination of rapid triggering and localisation
with pan-chromatic follow-up has provided a large and growing sample of burst
light-curves and spectra with which to confront theoretical models.  Using this
data, theoretical work continues to elucidate the progenitors of gamma-ray
bursts.  In this paper we focus on long gamma-ray bursts (LGRBs), for which the
model of \citet{Woosley93} has stood up well to the last two decades of
observations.  This posits the origin of LGRBs as accretion of a stellar-mass
quantity of material from an angular-momentum-supported disc around a
newly-formed stellar-mass black hole.  This model is supported by strong
observational evidence for an association between LGRBs and type Ib/c supernovae
\citep{Stanek03}; for a recent review of the evidence see \citet{HjorthBloom11}.
The specific rates of gamma-ray bursts, even once corrected for a narrow beaming
angle, are consistent with them forming from a small fraction of type Ib/c
supernovae \citep{Pod04}.

The main theoretical problem with this scenario is that a rapid rate of stellar
rotation is needed at the moment of core collapse, in order to endow the core
with sufficient angular momentum that its outer parts are unable to collapse
and instead form the accretion disc.  However, the presence of strong stellar
winds in the Wolf-Rayet stars that lead to type Ib/c supernovae would be expected to
brake the rotation of the stars and spin them down.
A number of possibilities have been considered to overcome this problem.  For
single stars, the most promising option is low metallicity, which then leads to
chemically homogeneous evolution of rapidly-rotating stars \citep[e.g.][]{Yoon05}.
The low metallicity suppresses stellar mass loss and thus prevents
wind-driven spin-down.  However, more recent observational evidence suggests
that a significant fraction of LGRBs form in higher-metallicity environments
where this mechanism is not plausible \citep[e.g.][]{Levesque10,Svensson12}.

Models proposed for the formation of gamma-ray bursts at higher metallicities
mostly focus on exploiting the reservoir of angular momentum stored in the
orbit of a binary stellar system in order to provide the spin of the exploding
star.  Here we focus on angular momentum transfer by tidal interaction
\citep{Pod04,Izzard04,ldk2006}.  In this scenario, tidal locking of the
progenitor core with a massive black-hole companion in the latest stages of its
evolution spins the core up to rapid rotation rates.  In
Section~\ref{sect:binpop} we use the \BSE{} stellar population synthesis code to
demonstrate the viability of this pathway and measure its frequency.  We then
analyse the effects of the binary companion on the accretion rate in the
subsequent LGRB, in an attempt to confront this model with the observational
data.  In Section~\ref{sect:model} we describe our model, and in
Section~\ref{sect:results} our results.  Section~\ref{sect:obscomp} contains a
comparison with observed long-duration gamma-ray bursts, and
Section~\ref{sect:sum} a summary.

\section{Binary population}
\label{sect:binpop}
To obtain a representative population of systems we synthesise a population of
BH--BH binaries, utilising the rapid binary population synthesis code \BSE{}
\citep{hurley2002} with the modifications of \citet{Church11} which for
convenience we summarise here.  We have
modified \BSE{} following \citet{bel2002b} and \citet{bel2008} to include a more
realistic prescription for compact object masses, hypercritical accretion during
common envelope evolution, and delayed dynamical instability in mass transfer
from helium stars in binaries with a large mass ratio.  We distribute stellar
masses according to the \citet{kroupa1993} IMF.  Binary semi-major axes are
chosen from a distribution flat in $\log a$ between 1 and $10^4\,\Ro$.  The
initial eccentricities of the binaries have
little effect on their evolution and are set equal to 0.1 for all binaries.
Neutron stars receive a natal kick chosen from the bimodal distribution of
\citet{acc02}.  We do not apply a kick to black holes upon formation.

\subsection{Criterion to obtain a tidal-spinup gamma-ray burst}
Following \citet{ldk2006}, we require that the material at the
edge of the stellar core must have sufficient angular momentum to fall back into
a disc at a radius outside the last stable orbit.  This means that
the specific angular momentum $j$ of material at the edge of a core of mass
$M_{\rm c}$ must satisfy $j>\sqrt{6}GM_{\rm c}/c$.  Taking the core to be
rotating as a solid body and be tidally locked to the binary, we apply Kepler's
law to obtain a critical orbital separation $a_{\rm crit}$ of
\begin{equation}
a_{\rm crit} = 7.36\,\Ro\,\left(\frac{M_{\rm
c}}{1.7\,\Mo}\right)^{-2/3}\left(\frac{M_{\rm T}}{20\,\Mo}\right)^{1/3},
\end{equation}
where we have taken a typical core radius for massive stars of $R_{\rm
c}=0.2\,\Ro$ and $M_{\rm T}$ is the total mass of the binary at this point.
For binaries that satisfy this criterion we assume that tidal locking takes
place at the point where the orbital separation is minimised.  This gives the
specific angular momentum of the material at the edge of the core as
\begin{equation}
j_{\rm shell} = 6.97\times10^6\,{\rm km^2\,s^{-1}}\,\sqrt{\frac{M_{\rm
T}}{20\,\Mo}}\left(\frac{a_{\rm min}}{3\,\Ro}\right)^{-3/2}.
\end{equation}

For cores that match this criterion, the material
further inside the core also acquires a substantial angular momentum.  The black
hole that forms from such a core will have a dimensionless spin parameter
close to unity.  This makes it a good candidate for powering a long gamma-ray
burst via the mechanism of \citet{BlandfordZnajek77}.

\subsection{Derived population}
\begin{figure}
\includegraphics[width=\columnwidth]{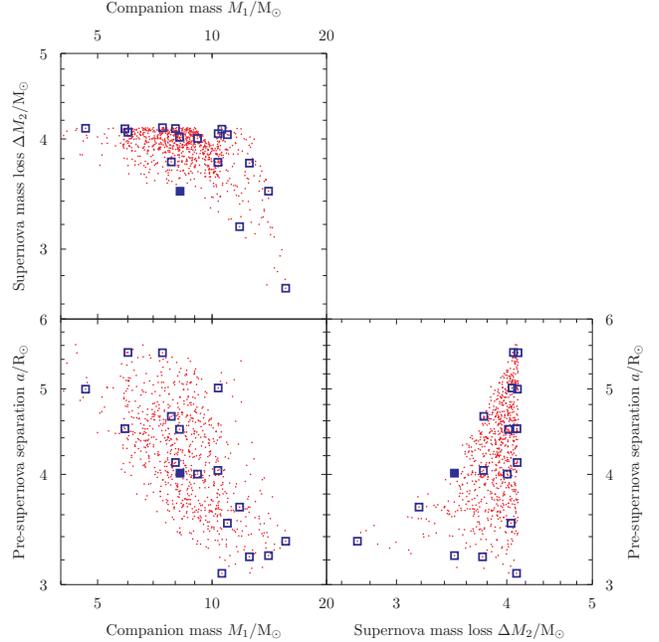}
\caption{The population of binaries that we expect to produce tidal spin-up
gamma-ray bursts.  In the bottom-left panel we plot the mass of the companion
black hole, $M_1$, against the binary separation before the second supernova, 
$a$.  The top-left panel shows the mass lost in the supernova, $\Delta M_2$, 
against $M_1$.  The lower-right panel shows $a$ against $\Delta M_2$.  In
BSE $\Delta M_2$ is a function of $M_2$ only.  Red dots show all the
binaries produced by our population synthesis.  Blue squares show the subset of
17 representative systems that we chose to investigate and the filled blue
square shows the location of our typical system.}
\label{fig:binpop}
\end{figure}

The population of binaries that we obtain from our \BSE{} runs is shown
in Figure~\ref{fig:binpop}.  We plot the relationships between the companion mass,
$M_1$, the pre-supernova binary semi-major axis, $a$, and the mass lost by the
exploding star, $\Delta M_2$.  The binaries where the criteria for a tidal-spinup LGRB
are satisfied are marked with red dots. We choose a sub-sample of 17 systems that
cover the parameter space evenly to investigate further; these are marked with
blue squares.

To calculate a representative rate for these systems we adopt the approach of
\citet{Pod04}, along with their rates for LGRBs and core-collapse supernovae in a
typical galaxy.  We present the rates for our binaries in Table~\ref{tab:rates}.
To calculate our rates we employ a \citet{kroupa1993} IMF and a minimum mass for
a core-collapse supernova of $8\,\Mo$.  Our conservative estimate takes a binary
fraction of 0.5 for all stars and assumes that their masses are drawn
independently from the IMF.  The more optimistic estimate -- which in our
judgement is more realistic -- takes the binary fraction to be unity for massive
stars and employs a distribution of mass ratios $f(q)=2q$, which biases the
sample towards equal-mass binaries.

\begin{table}
\begin{tabular}{ll}
\hline
Object & Rate \\
       & yr$^{-1}$/galaxy\\
\hline
Core-collapse supernovae & $7\times10^{-3}$\\
Gamma-ray bursts (beaming angle of $5^\circ$) & $3\times10^{-5}$ \\
\hline
Tidal spin-up LGRBs (conservative estimate)    & $2.6\times10^{-9}$\\
Tidal spin-up LGRBs (optimistic estimate)      & $1.2\times10^{-6}$\\
\hline
\end{tabular}
\caption{Estimated event rates for a typical galaxy.  Values for core-collapse
supernovae and gamma-ray bursts are taken from \citet{Pod04}.} 
\label{tab:rates}
\end{table}

Our rates, taken at face value, are quite small; even the optimistic estimate is
 lower than the LGRB rate.  This is relatively unsurprising; we require a
binary that forms two black holes, one of them rather massive, in a binary with
an unusual configuration.  It is perhaps unsurprising that such binaries are
relatively rare compared to, for example, NS--NS binaries.  However, as we show,
making a simple change to the mass-ratio distribution in the binary can make a
very large difference to the predicted rate.  We have not explored further any
of the uncertainties in our assumptions about the binary evolution and how they
might affect the population of binaries produced, but we are confident that,
were we to do so, we would be able to match the observed GRB rate using
reasonable assumptions.  For example, the number of sufficiently spun-up
binaries that we produce is rather sensitive to our assumptions about the radius
of the helium core at the closest point in the evolution.  We have also not
evaluated whether the injection of angular momentum into the falling-back
material by interaction with a companion can increase the population of binaries
with sufficient angular momentum to form a disc.  There are also
potentially other possible routes to making gamma-ray bursts via tidal spin-up
that we have not considered.  For example, \citet{MMendez11} show that the high
spin of black holes in two Galactic binaries is consistent with them being the
relics of gamma-ray bursts, as predicted by \citet{Lee02}.  Such binaries are
likely to be relatively common compared to the massive double black hole
binaries that we consider here.

\subsection{The effects of natal kicks}
\label{sect:kicks}
There is some evidence that black holes receive natal kicks \citep[e.g.][see
also Repetto et al. submitted]{Brandt95,Gualandris05}.\nocite{Repetto12}  To
investigate the likely effect of natal kicks on the processes that
we are modelling we consider the effect of a moderate kick -- $100\,{\rm
km\,s^{-1}}$ -- in the four directions in the orbital plane: $\pm \{x,y\}$.  We
do not consider kicks in the $z$ direction because this breaks the symmetry of
the problem about the orbital plane, and hence evaluating the behaviour of the
subsequent accretion flow is more complex.  In each case we assume that the
material that is ejected and subsequently falls back on to the black hole
receives the same kick as the newly-formed black hole itself.  This is the
expected behaviour if the kick is caused by a mechanism intrinsic to the
supernova explosion, which previous work has found to be the most likely case
for the Galactic black-hole binary Cygnus X-1 \citep{Axelsson11}.

\section{Modelling the fallback process}
\label{sect:model}

To treat the fallback process we utilise the simulations of \citet{mwh01}.  They
find a broadly self-similar behaviour of the accretion of material 
on to the nascent black hole,  with an early plateau phase followed by the
$t^{-5/3}$ decline of \citet{Chevalier89}.  We fit this with a two-piece
solution:
\begin{equation}
\dot{M}=
   \begin{cases}
        \kappa & 100\,{\rm s}<t<t_{\rm plateau} \\
        \kappa(t/t_{\rm plateau})^{-5/3} & t>t_{\rm plateau}
   \end{cases}
\label{eqn:mdot}
\end{equation}
where $\log_{10} (t_{\rm plateau}-100\,{\rm s}) = 2.2$; i.e. $t_{\rm plateau}\simeq
260\,{\rm s}$, and $\kappa$ is chosen to obtain the correct total fallback mass.  To
obtain the fallback mass we assume that the inner $2\,\Mo$ of the final black
hole mass is acquired during the initial core collapse; the remainder forms the
fallback mass.  This prescription leads to a fallback mass of between 1 and
$3\,\Mo$.

\subsection{Particle trajectories}

By using the results of hydrodynamic simulations of black-hole formation in
fallback supernovae in the form of Equation~\ref{eqn:mdot} we avoid having to
compute the hydrodynamics of the accretion shocks ourselves.  Instead we treat
our particles as interacting only under gravity.  We launch our particles with a
distribution of velocities, chosen so that, in the absence of a companion star,
the particles would fall back with times that reproduce the accretion rate given
by Equation~\ref{eqn:mdot}.  For purely radial motion the return time $t_{\rm
return}$ for a particle launched at radius $r_0$ from a star of mass $M_i$ is
given by
\begin{equation}
t_{\rm
return}=\sqrt{\frac{r_0^3}{GM_i}}\left(\frac{\alpha}{1-\alpha}\right)\left(\alpha +
             \sqrt{\frac{\alpha}{1-\alpha}}\tan^{-1}\sqrt{\frac{\alpha}{1-\alpha}}\right),
\label{eqn:tret}
\end{equation}
where $\alpha$ is related to the particle's launch velocity $v_0$ according to
\begin{equation}
v_0=\sqrt{\alpha \frac{2GM_i}{r_0}}.
\end{equation}
We invert Equation.~\ref{eqn:tret} numerically in order to produce a distribution of
$\alpha$ from the accretion rate of Equation.~\ref{eqn:mdot}.  Because the star is
rotating the initial velocities are not radial; however, because
$r_0$ is small compared to the maximum distance that the
particle reaches from the exploding star, Equation~\ref{eqn:tret} suffices.  We
choose the value of $r_0$ so that the rotational velocity at the equator is 20\%
of the escape velocity at that radius; thus even at launch the particle
velocities are predominantly radial.

We follow the motion of the particles under the gravitational force of the two
black holes.  As we are using Equation~\ref{eqn:mdot} to
account for the self-interaction of the gas we ignore the mass of the gas particles
and any non-gravitational forces; in practice, therefore, we solve the
reduced three-body problem numerically.  We use the Bulirsch-Stoer method as
given by \citet{nr} to integrate the orbits of the stars and particles.

We follow the motion of the particles until they cross the orbital plane ($z=0$)
for the first time.  The angular momentum vector of the stellar rotation is
parallel to that of the orbit as it is induced by tidal spin-up of the exploding
star: hence the system is symmetric under reflection in the orbital plane.
Because we can expect any given particle that crosses the orbital plane to have
a counterpart with the equal and opposite $z$--velocity a disc will form in the
plane.  We measure the mass and momentum flows into the disc by monitoring the
rate at which particles cross the plane and the specific angular momentum that
they possess when they do.  We assume that any particles that cross the plane
sufficiently close to a black hole to be within the last stable orbit are
accreted directly.

\subsection{Accretion disc}
Following \citet{Perna06} we use a simple treatment of the accretion disc based
on the viscous timescale.  We assume that a ring of material, represented by one
of our particles that has accreted into the $x-y$ plane, will move inward
through the disc via viscous dissipation of angular momentum.  The rate at which
a particle moves inward is given by
\begin{equation}
\frac{\ud r}{\ud t}=-\frac{r}{t_0},
\end{equation}
with characteristic viscous timescale $t_0$ given by
\begin{equation}
t_0=\left(\frac{R_{\rm disc}}{H_{\rm disc}}\right)^2\,\frac{1}{\alpha\Omega_{\rm
Kep}}
\end{equation}
for a disc of scale height $H$ and radius $R$.  The orbital angular frequency at
radius $r$ in the disc around the compact object with mass $M_i$ is given by
\begin{equation}
\Omega_{\rm Kep}=\sqrt{\frac{GM_i}{r^3}}.
\end{equation}
We take the ratio $R_{\rm disc}/H_{\rm disc}=10$ to be fixed
throughout the evolution.  Although the disc is probably less flat than this at
early times the viscous timescale then is short enough for the results to be
basically unaffected.  A particle unaffected by other
particles, arriving at the disc at time $\tau$, then spirals towards the central
black hole according to
\begin{equation}
r(t)=r(\tau)\left[1-\frac{3}{2}\frac{t-\tau}{t_0(\tau)}\right]^{2/3}.
\end{equation}
We set $r(\tau)$ to be the particle's circularisation radius; that is, we assume
that the material that it represents will self-collide to form a circular ring,
conserving specific angular momentum.  Thus the initial radius at which the
particle joins the disc is given, in terms of its specific angular momentum $j$,
by
\begin{equation}
r(\tau)=r_{\rm circ}=\frac{j^2}{GM_i}.
\end{equation}
In the process of circularising, i.e.~moving between the radius from the central
object at which it enters the plane and that at which it circularises, the
material may collide with gas already present in the disc.  If this happens we
merge the two colliding particles and calculate a new specific angular momentum,
and hence circularisation radius.  This process is repeated until the particle
has successfully circularised.  Particles that circularise outside the
instantaneous Roche lobe of either star are assumed to be lost from the system.
We have tested that the accretion history resulting from this treatment
converges as we increase the number of particles, and hence the mass resolution
of the simulation.

\section{Results}
\label{sect:results}
We present first the results for a typical binary, indicated by the
filled square in Figure~\ref{fig:binpop}.  It has a companion
black-hole mass of $M_1=8.24\,\Mo$; the newly-formed black hole has a total
mass, including the material falling back, of $M_2=4.29\,\Mo$, having lost
$\Delta M_2=3.49\,\Mo$ in the explosion.  The semi-major axis of the orbit at
the time of explosion is $a=4.01\,\Ro$.

\begin{figure}
\includegraphics[width=\columnwidth]{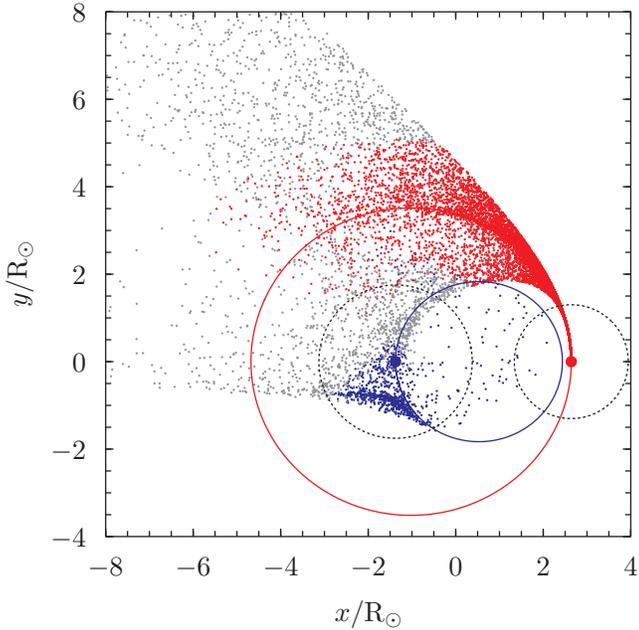}
\caption{The locations where particles cross the orbital plane in our reference
simulation.  The large red and blue points are the initial positions of the
exploding star and companion black hole respectively.  The red and blue solid lines
show the post-supernova orbits of the two stars.  Initially the exploding
(red) star is moving in the $+y$ direction and the companion black hole (blue)
in the $-y$ direction.  The red dots are particles that fall into a disc around
the newly-formed black hole; similarly, the blue dots are particles that fall
into a disc around the companion black hole.  Grey dots are particles that
fall further from either object than its Roche lobe radius and hence are
expected not to be accreted.  The black dashed lines show the initial radii of
the Roche lobes around the two stars.  The origin of the co-ordinate system is
the centre of mass of the post-explosion system.  For clarity only 1\% of the
particles used in the final simulation are plotted.} 
\label{fig:xyplane}
\end{figure}

\begin{figure}
\includegraphics[width=\columnwidth]{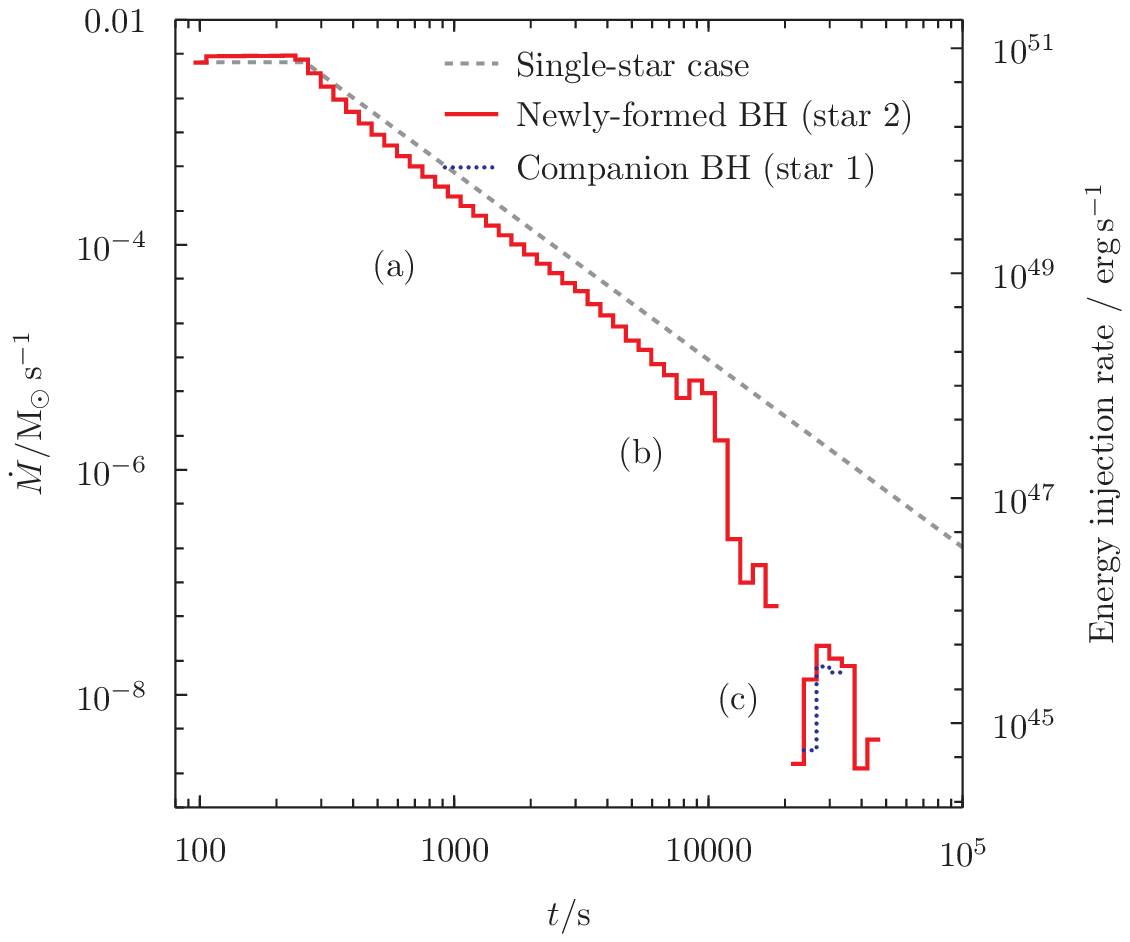}
\caption{The accretion rate on to the black holes in our typical example system.
The grey dashed line shows the accretion rate of Equation~\ref{eqn:mdot}, which
would be obtained in the absence of a companion.
The red solid line shows accretion on to the newly-formed black hole (star 2) whilst
the blue dotted line is for the companion (star 1).  The plot includes the
effect of the accretion disc and the Roche lobes of the two stars.  Of the
$2.29\,\Mo$ of material in the fallback shell $1.90\,\Mo$ is accreted on to star
2 and $0.02\,\Mo$ on to star 1; the remaining $0.37\,\Mo$ is lost from the
system.  Characteristic features are (a) the initial plateau and roughly
$t^{-5/3}$ decline phases, (b) the break in the accretion rate at roughly
$10^4\,{\rm s}$ followed by a steep decline, and (c) some late-time accretion
activity (at around $3\times10^4\,{\rm s}$).  The right-hand axis shows the
energy injection from accretion, assuming that the efficiency of conversion of
mass to energy is $\eta=0.1$.
}
\label{fig:accrRate}
\end{figure}

In Figure~\ref{fig:xyplane} we plot the location of each particle in the
simulation when it crosses the orbital plane.  The exploding star is initially
moving in the $+y$ direction.  As it proceeds round its orbit, the particles
that fall back towards it have been affected to a greater degree by interaction
with the companion, so their deviation from the stellar position is increased.
These particles typically also have more specific angular
momentum and hence enter the accretion disc at larger radii.  After a
some time -- about one quarter of the post-explosion orbital period of the
binary -- the particles start to mostly fall back outside the Roche lobe of the
star and hence the mass inflow rate into the disc reduces sharply.  Some
particles instead, however, fall into the Roche lobe of the companion black hole
(blue dots in Figure~\ref{fig:xyplane}).  The accretion rate on to the two stars
as a function of time in this typical model is shown in
Figure~\ref{fig:accrRate}.

Our typical model shows a number of features that are common to the models that
we have run.  There is an initial plateau followed by a declining power-law
accretion rate.  This follows directly from our input accretion rate; the
particles falling back at early times are largely unaffected by the presence of
the companion.  At roughly $10^4\,{\rm s}$ there is a break in the accretion
curve, which transitions to a steeper power law.  Finally there is a period of
zero accretion at about $1.75\times10^4\,{\rm s}$, followed by a flare that
lasts for several $10^4\,{\rm s}$.  During this flare there is additionally some
accretion on to the companion black hole.

\subsection{Fits to the accretion behaviour}

\begin{figure}
\includegraphics[width=\columnwidth]{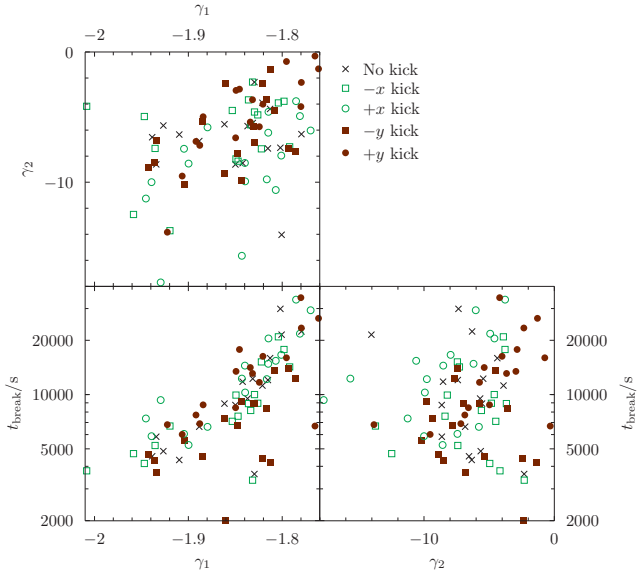}
\caption{Correlations between different parameters in our power-law fits to the
accretion rates on to the two black holes.  The power-law applicable to the decay
between $200\,\s$ and the break time $t_{\rm break}$ is denoted by
$\gamma_1$, whereas the power law between $t_{\rm break}$ and the first
cessation of accretion is denoted by $\gamma_2$.  The bottom-left panel
shows the correlation between $\gamma_1$ and $t_{\rm break}$, the top-left
panel that between $\gamma_1$ and $\gamma_2$, and the bottom-right panel that
between $\gamma_2$ and $t_{\rm break}$.  Black crosses are for models without
natal kicks, brown filled circles and squares are models with kicks in the $+y$
and $-y$ directions, and green open circles and squares are models with kicks in
the $+x$ and $-x$ directions.  Only the first of these three shows a significant
correlation.
}
\label{fig:powerLawFits}
\end{figure}

To gain some insight into the implications of our model we fit each of our 17
characteristic systems with a broken power-low model of the form
\begin{equation}
\dot{M}\propto
   \begin{cases}
        t^{\gamma_1}  & 200\,{\rm s}<t<t_{\rm break} \\
        t^{\gamma_2}  & t_{\rm break}<t<t_0,
   \end{cases}
\label{eqn:broken}
\end{equation}
where $\gamma_1$, $\gamma_2$ and $t_{\rm break}$ are obtained by fitting.  The
time at which the accretion first ceases, $t_0$, is obtained from the accretion
curve.  In Figure~\ref{fig:powerLawFits} we present the results of the fits.
The bottom-left panel shows the break time, $t_{\rm break}$, plotted as a
function of $\gamma_1$, the power-law index of the portion of the accretion
curve before the break.  The plot shows that the accretion curves are somewhat
steeper than the $t^{-5/3}$ input power-law; shorter break times are correlated
with more steeply declining accretion.  The power-law index of the second,
post-break power law segment is much steeper, typically between -5 and -10, and
does not show any correlation with the other parameters.

\begin{figure}
\includegraphics[width=\columnwidth]{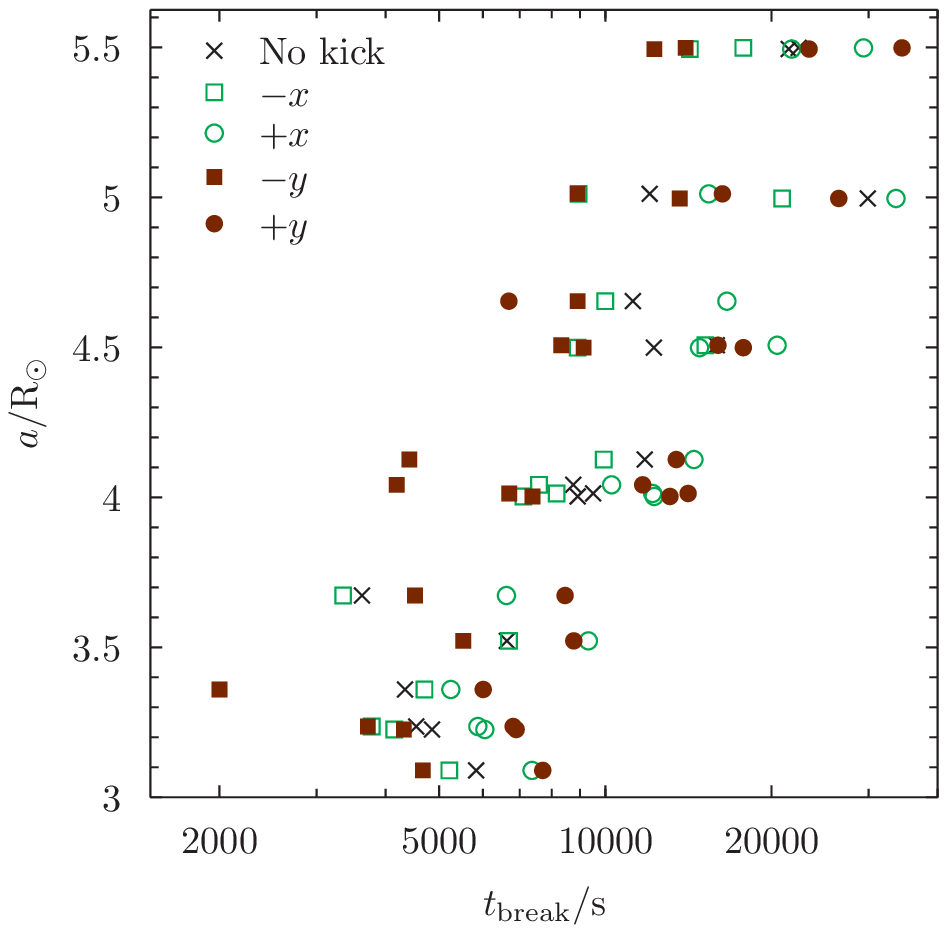}
\caption{The correlation between the binary semi-major axis $a$
and the break time $t_{\rm break}$.  Wider binaries only affect particles
falling back at later times, thus making $t_{\rm break}$ later for larger
semi-major axis.  
Black crosses are for models without
natal kicks, brown filled circles and squares are models with kicks in the $+y$
and $-y$ directions, and green open circles and squares are models with kicks in
the $+x$ and $-x$ directions.  
}
\label{fig:correlation}
\end{figure}

In Figure~\ref{fig:correlation} we present the correlation between the
semi-major axis of the binary, $a$, and the accretion break time.  The evident
correlation has a simple explanation; a wider orbit means that the material has
to travel further before it interacts significantly with the companion black
hole, increasing $t_{\rm break}$.  This in turn also means that the angular
momentum imparted to the material which is accreted before the break is
lessened; hence there is a positive correlation between $a$ and $\gamma_1$.


In both Figures~\ref{fig:powerLawFits} and \ref{fig:correlation} we plot both
systems computed under the assumption of black-hole kicks, as well as unkicked
systems.  We find that the behaviour of the kicked and unkicked systems is
broadly similar to the unkicked case, and there are no marked trends between the
different kicks.  This suggests that a moderate kick is unlikely to have a
strong effect on the fallback process.


\section{Comparison with observations}
\label{sect:obscomp}

It is difficult to compare our results directly with observations.  In order to
go from the black hole's accretion history to the observed photometry and
spectral features it is necessary to model the propagation of the fireball and
jet through the exploding star, a highly involved procedure.  What we can say,
however, is that we expect to see evidence of continued energy injection during
the first approximately $10^4\,\s$ in the rest frame of the accreting black
hole.  We predict that the rate of this energy injection would decline quite
strongly, although we should caution against the use of our fitted values of
$\gamma$ as the decline of the injected luminosity.  As the system evolves and
the accretion rate decreases the accretion morphology will become more disc-like
and less advection dominated.  This will increase the fraction of the accretion
luminosity available to power a jet at later times.

Secondly, we predict a sharp reduction in the accretion luminosity at times of
around $10^4\,{\rm s}$, which we would expect to appear as a break in the light
curve.  These times are consistent with typical times found by \citet{Evans09}
for the end of the plateau phase in canonical light curves, whose distribution
also peaks at about $10^4\,{\rm s}$.  Light curves that show a single,
steepening break also have break times around $10^4\,\s$, although the
distribution is much wider.  Our break times are also consistent with the subset
of the sample of light curves presented by \citet{Li2012} that show shallow
decay features.  Interpretation of the optical light curves is complicated,
however, by spectral breaks which can occur at relevantly late times.  

The physical origin of the break in our models is the truncation of accretion by
the Roche lobe of the exploding star.  Material that falls outside the Roche
lobe is not bound to either star, and hence will be lost from the system.  As
time progresses, the material falling back has traversed a larger distance from
the exploding star and hence is more affected by the gravitational force of the
companion.  Thus it falls back at a greater distance from the exploding star.
The break time can be understood as the time at which this distance is the Roche
lobe radius of the star.  This is the origin of the correlation visible in
Figure~\ref{fig:correlation}, as in wider binaries the Roche lobe is larger and
thus the break time is later.

\subsection{Late-time highly-variable accretion}
\begin{figure}
\includegraphics[width=\columnwidth]{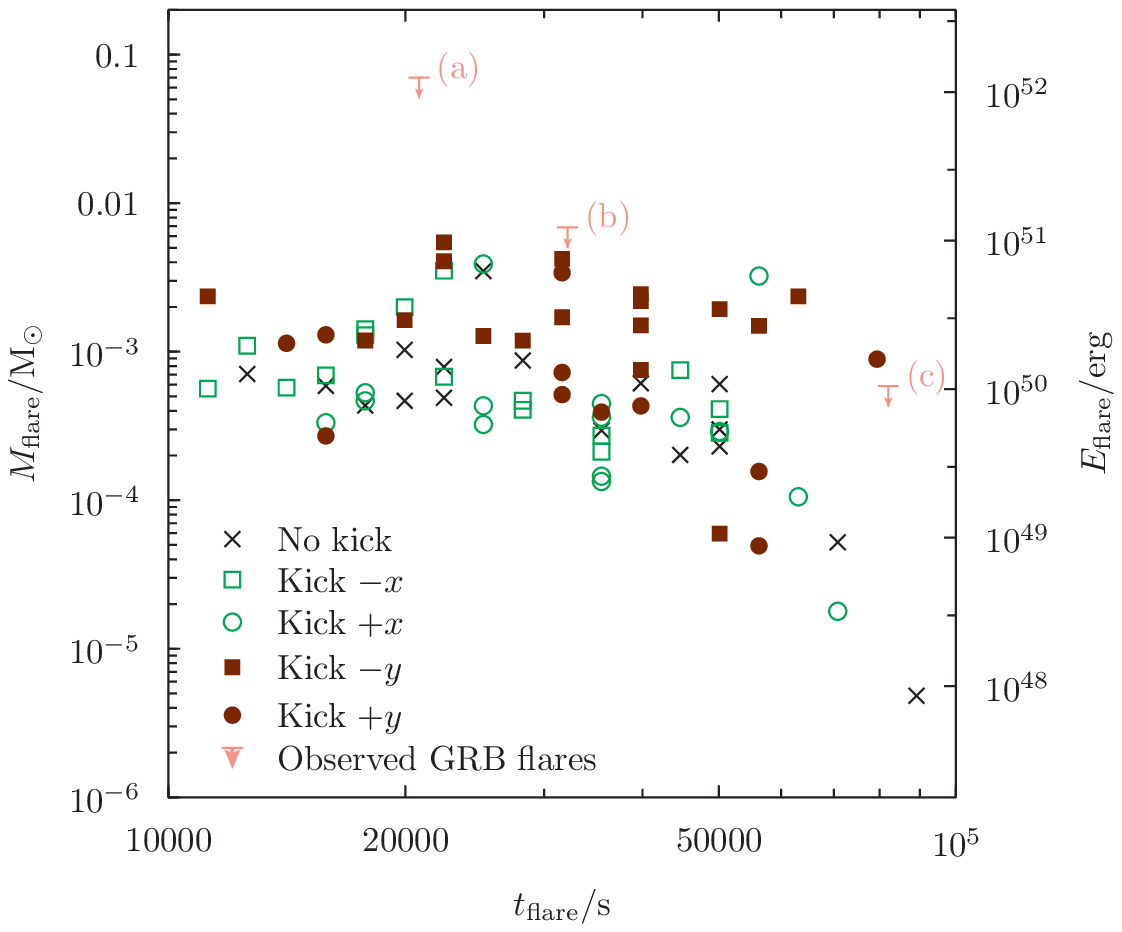}
\caption{Episodes of highly-variable late-time accretion in our models.  We plot
the total mass accreted in late-time flares, $M_{\rm flare}$, as a function of
the break time $t_{\rm break}$.  As in previous figures, black
points are for models without natal kicks, brown filled circles and squares are
models with kicks in the $+y$ and $-y$ directions, and green open 
circles squares are are models with kicks in the $+x$ and $-x$ directions.  
Flare energies (right-hand axis) are calculated assuming mass
is converted into energy with an efficiency of $\eta=0.1$.  The pink upper
limits represent the energy in late-time flares from three long-duration gamma-ray
bursts, (a) GRB~050502B, (b) GRB~070107 and (c) GRB~070318.  In calculating
these points we assume that the energy emission at  the time of the flare is
isotropic.
}
\label{fig:flares2} 
\end{figure}

The majority of our models include some form of late-time highly-variable
accretion.  This typically occurs at a few times $t_{\rm break}$, after the
second power-law component has ended.  In our typical system, for example, after
the main accretion episode has ended at around $2\times10^4\,{\rm s}$ there is a
renewed episode of accretion peaking at around $3\times10^4\,{\rm s}$.  Such
activity might power late-time flares.  In Figure~\ref{fig:flares2} we show the
properties of the flares in those of our models that show them.  We assume that,
once the accretion energy conversion efficiency has been taking into account,
all the energy liberated goes into the flare.  Having done this, the luminosity
generated is very comparable to that seen in late-time flares observed in the
X-ray and optical.  For comparison, we plot the energy in the late-time flares
of three observed long gamma-ray bursts, GRB~050502B, GRB~070107 and GRB~070318.
This is the subset of the sample considered by \citet{Curran08} that have
reasonably constrained photometric or spectroscopic redshifts and which are long
bursts.  We take the light curves for these flares from \citet{Evans09} and
redshifts from \citet{Xiao11}.  Although the observed flares appear to be
somewhat more energetic than our predictions, it is only necessary for them to
be beamed to one tenth of the sky at the time of the flare to be consistent with
our predicted values.  We re-produce the observed decline in flare energy
with time, which is also seen in optical data \citep{Li2012}.

\section{Discussion}
\label{sect:sum}

We have considered tidal spin-up compact binaries as a possible source for
long-duration gamma-ray bursts.  We show that, with appropriate assumptions, a
population synthesis approach predicts that these binaries will be produced.
Furthermore we show that, given two different -- but plausible -- assumptions
about the distribution of mass-ratios of high-mass binaries we predict two very
different formation rates.  Hence we suggest that the power of population
synthesis calculations to constrain progenitor models by analysis of expected
rates is rather weak, as argued by \citet{Tout05} for the similar problem of
Type Ia supernova progenitors.

A potential means of distinguishing this progenitor scenario, however, is that
in the case of a tidally spun-up progenitor, the star exploding as a type Ib/c
supernova has a black-hole companion.  We have modelled the effect of the companion on
the fallback process to investigate what observable consequences it would have
for the gamma-ray burst.  We conclude that our models possess the following
general features:

\begin{itemize}
\item Before about $10^4\,\s$, an accretion rate onto the newly-formed black
hole that declines as $t^{-1.85\pm0.05}$,
\item a break in the accretion rate at around $10^4\,\s$, with a steeply
declining accretion rate thereafter, and
\item in the majority of cases, highly-variable late-time accretion, leading to
flares at times between $10^4$ and $10^5\,\s$, of total energies around
$10^{50}\,{\rm erg}$.
\end{itemize}

These general features are broadly consistent with the observations of
long-duration gamma-ray bursts.  Bursts with a canonical-type light curve show
evidence of energy injection up until roughly the times that we predict, with a
break thereafter.  For the three bursts from the sample of \citet{Curran08} that
have a well-defined redshift and late-time bursts, the properties of the bursts
agree with our predictions, assuming a reasonable beaming fraction.  We
reproduce the tentative correlation between the time of the flare and the
energy.  We caution, however, that our model on its own cannot explain flares
that occur prior to $10^4\,\s$.  It does not preclude such flares from occurring;
however, they must have a separate origin, such as instabilities in the
accretion flow onto the newly-formed black hole.

Although we have only considered tidal spin-up by a massive compact companion,
it is also possible that the exploding star could be spun up by a main-sequence
companion.  \citet{Lee02} consider the evolution of the progenitors of low-mass
X-ray binaries.  They show that tidal forces from a low-mass main-sequence
companion and frictional effects during common-envelope evolution can be
sufficient to spin up the outer layers of the helium star, which subsequently
collapses to give a black hole.  We expect gamma-ray bursts produced by such a
system to also show breaks and flares similar to those described in this paper;
the binary has a similar semi-major axis and the same processes will operate.
We intend to undertake a quantitative study of these binaries in a subsequent
work.  

Finally, we investigate the effect of black-hole kicks on our scenario and show
that a moderate kick is expected to have a relatively small effect on the
observed properties of the burst.  We caution, however, that our model can only
encompass kicks that are directed within the orbital plane of the binary. Kicks
with a component parallel to the angular momentum vector would disturb the
geometry of the situation; this could make the accretion flow more disc-like and
less spherical, which we would expect to have significant consequences for the
baryon loading of the resulting burst.

\section*{Acknowledgements}
The authors would like to thank Johan Fynbo for useful discussions.  RPC is
funded by a Marie-Curie Intra-European Fellowship, grant No.~252431, under the
European Commission's FP7 framework.  CK acknowledges a Marie-Curie
International Incoming Fellowship under the European Commission's FP7
framework.  This work was supported by the Swedish Research Council (grants
2008-4089 and 2011-3991).  The calculations presented in this paper were
carried out using computer hardware purchased with grants from the Royal
Fysiographic Society of Lund.  This work made use of data supplied by the UK
Swift Science Data Centre at the University of Leicester.


\bibliographystyle{mn2e.bst}
\bibliography{new.bib}

\label{lastpage}

\end{document}